\RequirePackage{fix-cm}
\documentclass{svjour3}                     
\smartqed  
\usepackage{graphicx}
\usepackage{comment}
\usepackage{mathptmx}      
\usepackage{latexsym}
\usepackage{amsmath}
\usepackage{amssymb}
\usepackage{xcolor}
\usepackage{graphicx}
\usepackage{lipsum}

\newcommand{\kepler}{\emph{Kepler}}

\journalname{Celestial Mechanics and Dynamical Astronomy}
\begin{document}

\title{Tidal Decay and Stable Roche-Lobe Overflow of Short-Period Gaseous Exoplanets
}
%


\titlerunning{Tidal Decay and Overflow of Gaseous Exoplanets}        

\author{Brian Jackson		\and
        Emily Jensen		\and
        Sarah Peacock		\and
        Phil Arras			\and
        Kaloyan Penev
}


\institute{Brian Jackson \at
		Boise State University, Dept. of Physics,
		1910 University Drive, Boise ID 83725 USA\\
        \email{bjackson@boisestate.edu}        
        \and
		Emily Jensen \at
		Boise State University, Dept.\ of Physics,
		1910 University Drive, Boise ID 83725 USA
        \and
		Sarah Peacock \at
		Lunar and Planetary Laboratory, University of Arizona,
		1629 E University Blvd, Tucson, AZ 85721-0092
		\and
		Phil Arras \at
		Department of Astronomy, University of Virginia,
        Charlottesville, VA 22904-4325, USA
		\and
		Kaloyan Penev \at
		Department of Astrophysical Sciences, Princeton University, NJ 08544, USA
}

\date{Received: date / Accepted: date}

\maketitle

\begin{abstract}
Many gaseous exoplanets in short-period orbits are on the verge or are in the process of Roche-lobe overflow (RLO). Moreover, orbital stability analysis shows tides can drive many hot Jupiters to spiral inevitably toward their host stars. Thus, the coupled processes of orbital evolution and RLO likely shape the observed distribution of close-in exoplanets and may even be responsible for producing some of the short-period rocky planets. However, the exact outcome for an overflowing planet depends on its internal response to mass loss, and the accompanying orbital evolution can act to enhance or inhibit RLO. In this study, we apply the fully-featured and robust Modules for Experiments in Stellar Astrophysics (MESA) suite to model RLO of short-period gaseous planets. We show that, although the detailed evolution may depend on several properties of the planetary system, it is largely determined by the core mass of the overflowing gas giant. In particular, we find that the orbital expansion that accompanies RLO often stops and reverses at a specific maximum period that depends on the core mass. We suggest that RLO may often strand the remnant of a gas giant near this orbital period, which provides an observational prediction that can corroborate the hypothesis that short-period gas giants undergo RLO. We conduct a preliminary comparison of this prediction to the observed population of small, short-period planets and find some planets in orbits that may be consistent with this picture. To the extent that we can establish some short-period planets are indeed the remnants of gas giants, that population can elucidate the properties of gas giant cores, the properties of which remain largely unconstrained.
\end{abstract}

\section{Introduction}
\label{sec:introduction}

\begin{figure}
\includegraphics[width=\textwidth]{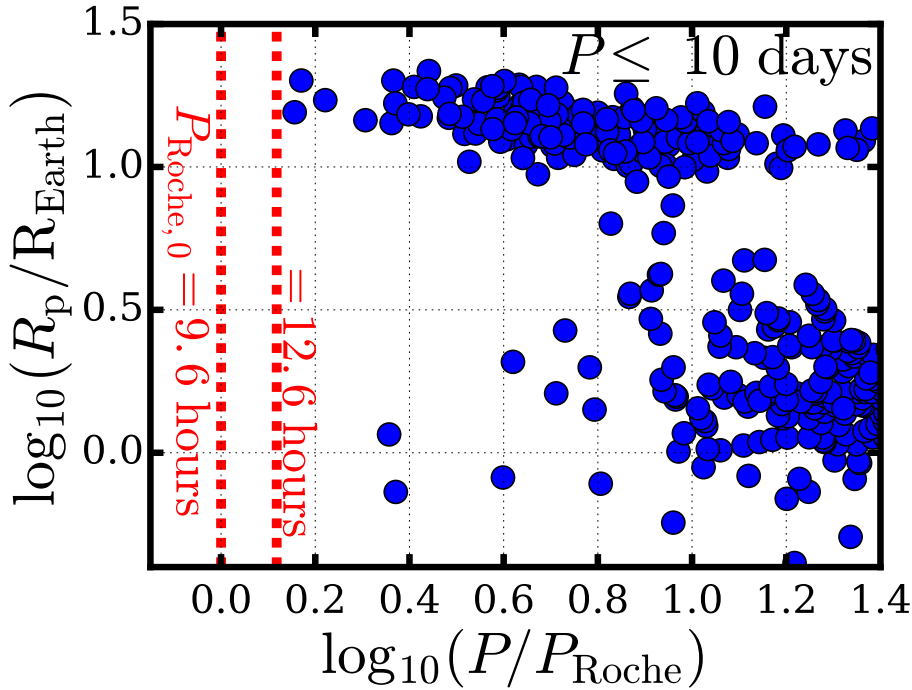}
\caption{Among planets with orbital periods $P \le $ 10 days, planetary radius $R_{\rm p}$ in Earth radii ($R_{\rm Earth}$) vs. the ratio of $P$ to the planet's Roche period $P_{\rm Roche}$, with $P_{\rm Roche, 0} =$ 9.6 hours for planets comprised of highly compressible fluid (also shown with the leftmost dashed, vertical line). The rightmost vertical line shows $P_{\rm Roche, 0} =$ 12.6 hours (for planets comprised of incompressible fluid with negligible bulk tensile strength). Data harvested from exoplanets.org on 2015 Dec 7.}
\label{fig:P-PRoche}
\end{figure}

From wispy gas giants on the verge of Roche-lobe overflow (RLO) to tiny rocky bodies already falling apart, extrasolar (or exo-) planets with orbital periods of several days and less challenge theories of planet formation and evolution. Although they are statistically rare, most current survey techniques favor their detection, and the population of known short-period planets has grown dramatically over the years. For example, among stars targeted by the \kepler\ Mission, \cite{Howard2010Occurrence} estimated 3.4$\pm$0.3\% of the GK dwarfs host planets with radii between 2 and 32 times the Earth's ${\rm R_{Earth}}$, and periods $P <$ 10 days, while almost 30\% of all \kepler's candidates lie within those ranges. 

Figure \ref{fig:P-PRoche} shows the radii $R_{\rm p}$ and orbital periods $P$ of short-period planets and how close they are to their Roche limits. \cite{Rappaport2013Roche} cast the classic Roche limit for a low-mass companion as a period $P_{\rm Roche}$ that only depends on a planet's density $\rho_{\rm p}$: $P_{\rm Roche} = P_{\rm Roche, 0} \left( 1\ {\rm g\ cm^{-3}} / \rho_{\rm p} \right)^{1/2}$. For planets comprised of highly compressible fluid, $P_{\rm Roche, 0} =$ 9.6 hours, while for planets comprised of incompressible fluid with negligible bulk tensile strength, $P_{\rm Roche, 0} =$ 12.6 hours. The x-axis of Figure \ref{fig:P-PRoche} assumes $P_{\rm Roche, 0} =$ 9.6 hours for all planets. \cite{Rappaport2013Roche} explored deviations from the simple $P_{\rm Roche}$ expression used here. 

The distribution of orbits for hot Jupiters extends to very near the planets' Roche limits, indicating that at least some hot Jupiters are on the verge of RLO. Others may actually be in the process of RLO. For example, \cite{2010Natur.463.1054L} pointed out that WASP-12 b, a hot Jupiter in a 19-hour orbit, may be undergoing RLO since its hot atmosphere likely extends up to planet's Roche lobe, even though the photosphere does not. 

Since the vast majority of short-period planets have circular orbits and probably synchronized rotation, tides raised on the planets by the stars have no effect on the orbital evolution, but for stars that are not tidally locked to their planets, tides raised on the stars by the planets can drive orbital evolution, long after eccentricities drop to zero. In cases where the host star rotates more slowly than the planet revolves, tides raised on the star lag the planet, and the resulting gravitational pull on the planet of this bulge reduces mechanical energy and transfers angular momentum from the orbit to the star's rotation, driving orbital decay \cite{2008CeMDA.101..171F}. In fact, the vast majority of known planet-hosting stars rotate more slowly than their close-in planetary companion and so fall into this category. An interesting exception is the $\tau$ Boo system, where the star may be tidally locked to the planet \cite{2008A&A...482..691W}. The HAT-P-11 system is also noteworthy for a possible 6:1 commensurability between the orbital and stellar rotation periods, as discussed in \cite{2014ApJ...788....1B}.

Energy and angular momentum considerations of tidal interaction indicate that many hot Jupiter systems in circular orbits are formally unstable against tidal decay \cite{1973ApJ...180..307C}. \cite{2009ApJ...692L...9L} pointed out that most hot Jupiter systems known at that time have insufficient angular momentum to reach a stable tidal equilibrium, although more recently, \cite{2015MNRAS.446.3676A} looked again and decided that many, but not most, hot Jupiters are unstable. Thus, at least some fraction of hot Jupiter systems will inevitably spiral toward their host stars. 

Whether the systems will spiral inward in less than the main sequence lifetimes of the host stars is not clear, however. The rate of in-spiral depends in part on the efficiency with which tidal energy is dissipated within the host stars, often parameterized by an efficiency parameter $Q_\star$, and the dissipation processes within stars that set $Q_\star$ are not well-understood. Theoretical studies make a variety of predictions regarding the nature and efficiency of tidal dissipation within stars. \cite{2016ApJ...816...18E} predicts a rich but complex dependence of dissipation on the tidal frequencies, and much work, including \cite{2007ApJ...655.1166P}, suggests dissipation primarily occurs within a star's convective zone. Therefore, stars with deeper convective zones (i.e., cooler stars) should exhibit greater tidal dissipation efficiency. 

Estimates for effective $Q_\star$ derive from this wide body of work and range from $10^5$ \cite{1996Natur.380..606L,2016ApJ...816...18E} up to $10^9$ and larger \cite{2011ApJ...731...67P}. Larger $Q_\star$-values correspond to slower tidal evolution rates, and assuming frequency-independence \cite{1966Icar....5..375G}, this range of $Q_\star$ indicates tides would take between 2 Myrs and 20 Gyrs to drive a Jupiter-like planet around a Sun-like star from a period of 1 day to its Roche limit ($\sim$ 8 hours).

We can appeal to additional observations for constraints on the degree of tidal evolution among short-period planets. \cite{2013ApJ...775L..11M} found that, among the many \kepler\ targets for which rotation periods have been estimated, short-period planets are less commonly observed transiting the most rapidly rotating stars. This observation is qualitatively consistent with the idea that planets that have been accreted by their star via tidal decay deposited significant angular momentum in the host stars' outer envelope, giving them unusually high rotation rates \cite{2009ApJ...698.1357J}. 

\cite{2014A&A...565L...1P} sought the signature of tidal interactions by comparing members of stellar binaries in which one of the stars hosts a short-period planet and the other does not. They estimated X-ray activity (which scales with stellar rotation rate) for several widely separated binary stars and found that those stars with relatively deep convective zones exhibited enhanced X-ray activity, and hence more rapid rotation, than expected based on the X-ray activity of the partner stars without planets. \cite{2015ApJ...807...78F} studied some of these systems and found that the enhanced rotation rates for the planet-hosting stars pointed to substantial tidal decay of the accompanying planets. 

\begin{figure}
\includegraphics[width=\textwidth]{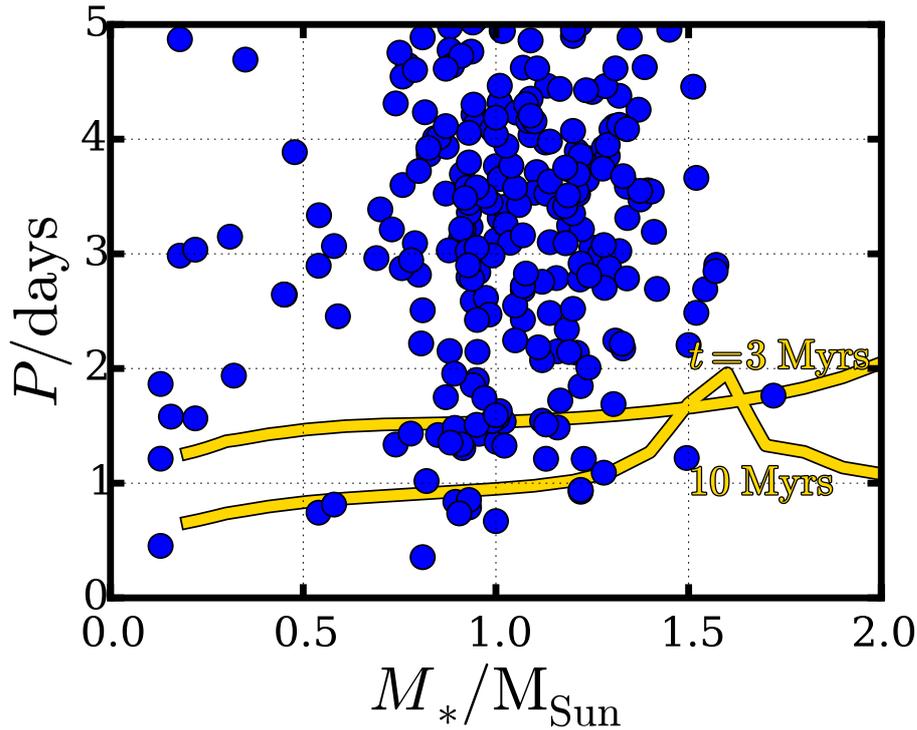}
\caption{Orbital periods $P$ vs. host star mass $M_\star$ for many short-period planets. The yellow lines show the periods corresponding to the stellar surface at 3 and 10 Myrs after stellar formation as modeled using MESA \cite{2011ApJS..192....3P}. The data for this plot were harvested from exoplanets.org on 2015 Jul 8.}
\label{fig:P-vs-Mstar}
\end{figure}

The period distribution of planets very close to their host stars may also point to non-negligible tidal decay. Figure \ref{fig:P-vs-Mstar} plots the periods for many short-period planets against the masses estimated for their host stars. Also shown as yellow lines are the periods corresponding to the stellar surfaces at 3 and 10 Myrs after formation, as modeled by the Modules for Experiments in Stellar Astrophysics suite (MESA) \cite{2011ApJS..192....3P}. One hypothesis for the origins of short-period planets involves migration while the planets are embedded within the protoplanetary gas disk \cite{2009AREPS..37..321C}, and so that evolution must have occurred within the disk lifetime. Disk lifetimes are thought to be less than 10 Myrs \cite{2013MNRAS.434..806B}. Many of the planets shown actually lie within the region occupied by their host star at such early times, which argues that the planets arrived at their orbits long afterward. Late-stage tidal decay of their orbits is an obvious explanation, although that evolution may have resulted from excitation of orbital eccentricity by additional planets and tidal dissipation within the planet \cite{2014PhDT.........4V} and not within the star. 

\begin{figure}
\includegraphics[width=\textwidth]{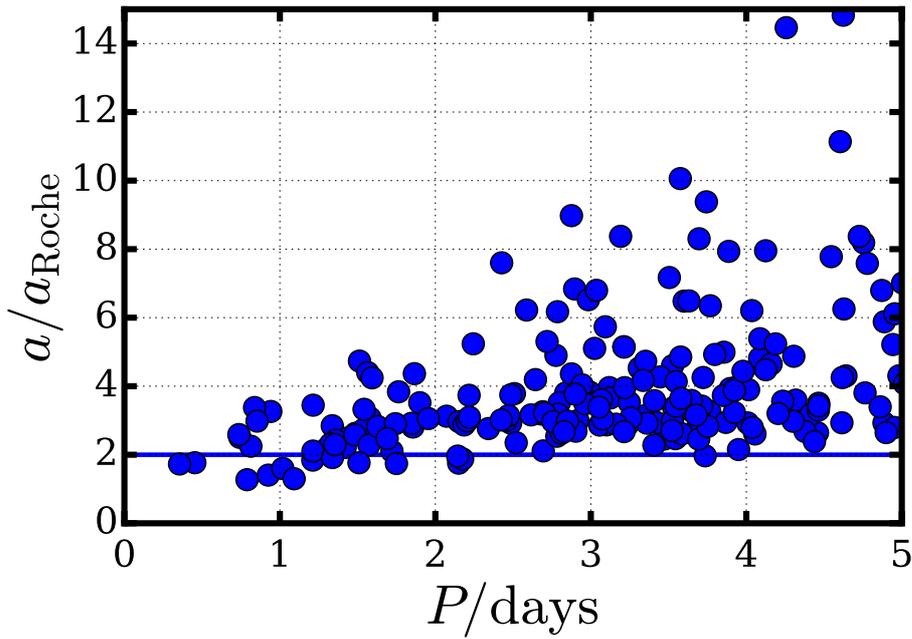}
\caption{The blue circles represent the ratio of a planet's orbital semi-major axis $a$ to its Roche limit $a_{\rm Roche}$. The horizontal blue lines shows the expected minimum ratio for planets having become close-in via circularization of a highly eccentric orbit \cite{2006ApJ...638L..45F}. The data for this plot were harvested from exoplanets.org on 2015 Jul 8.}
\label{fig:a-aRoche}
\end{figure}

An alternative origin scenario involves the dynamical excitation of a planet's orbital eccentricity to values near unity, giving a planet formed in a more distant orbit a pericenter distance close enough that tides can circularize and shrink the orbit \cite{2007ApJ...669.1298F,2011ApJ...735..109W}. The smallest pericenter allowed for such a planet is its Roche limit, and so \cite{2006ApJ...638L..45F} pointed out that, since tidal damping within the planet should nearly conserve the orbital angular momentum, the smallest semi-major axis $a$ at which the orbits would circularize is twice the planet's Roche limit. Using the same population as in Figure \ref{fig:P-vs-Mstar}, Figure \ref{fig:a-aRoche} shows each planet's $a$ and the semi-major axis for its Roche limit $a_{\rm Roche}$. ($a_{\rm Roche}$ was estimated from the available system parameters using data harvested from exoplanets.org on 2015 Jul 8.) The vast majority of planets lie outside $2\times a_{\rm Roche}$, but a small cluster of the short-period planets actually lies within. If those planets originated in the way described here, then their presence interior to $2\times a_{\rm Roche}$ can be explained by tidal decay of their orbits subsequent to their arrival into short-period orbits.

This body of circumstantial evidence suggests at least some close-in exoplanets undergo significant orbital decay and eventually mass transfer to the host star once they encounter their Roche limits, but what happens to the overflowing planets? How long does the RLO take, and are the remnants hidden amongst the currently observed population of close-in planets? Do the remnants have physical or orbital properties that could distinguish them? 

As for binary star systems, the timescales of RLO and accompanying orbital evolution depend very sensitively on the mass-radius relationship for the planets undergoing RLO \cite{Rappaport1982}. Recently, \cite{2014ApJ...793L...3V} and \cite{2015ApJ...813..101V} applied state-of-the-art planetary mass-radius relationships to investigate the outcomes of mass transfer for hot Jupiter systems for a range of assumptions about the details of the mass transfer. As we explain in Section \ref{sec:dynamics_of_planetary_rochelobe_overflow}, those studies found mass transfer can drive significant orbital expansion for a planet overflowing its Roche lobe, often stranding a remnant planetary core in an orbital period of several days, far from the Roche limit for the progenitor planet. The latter study also found an anti-correlation between the remnant mass and orbital period and argued that many of the small planetary objects recently discovered in ultra-short periods ($<$ 1 day) \cite{2014ApJ...787...47S,2016arXiv160306488A} are too close-in to be the remnants of hot Jupiters.

In this study, we re-visit the planetary mass transfer process, building on those previous studies to explore the orbital periods expected for the remnants of hot Jupiters. We also suggest that the remnants may appear as gas-rich super-Earth/sub-Neptunes with orbital periods that depend sensitively on the mass of the planet's solid core and with unusually low-density atmospheres. Our goal here is to develop specific theoretical predictions that can then be compared in detail to observations. We make a preliminary attempt at this comparison as well.

We focus here on stable mass transfer, which requires specific relationships between the details of the mass transfer and the response of the overflowing planet to mass loss. Unstable mass transfer involves rapid disruption and accretion of a planet on timescales comparable to the orbital period \cite{2002ApJ...565.1107P} and may also be important in some cases. We briefly discuss the conditions required for stable transfer in Section \ref{sec:dynamics_of_planetary_rochelobe_overflow} and leave unstable transfer for future work.

The plan for this paper is as follows: In Section \ref{sec:dynamics_of_planetary_rochelobe_overflow}, we review the dynamics of mass transfer or Roche-lobe overflow, with approximations tailored for planetary systems. In Section \ref{sec:RLO_Results_from_MESA}, we present the results of a large suite of evolution calculations using MESA. In Section \ref{sec:Evolution_of_the_Roche_Limit_for_a_Disrupting_Gaseous_Planet}, we explore a simple relationship between the core mass for a remnant and its orbital period and compare that prediction to observations. Finally, in Section \ref{sec:Discussion_and_Conclusions}, we conclude and discuss future work.

\section{Dynamics of Planetary Roche-Lobe Overflow}
\label{sec:dynamics_of_planetary_rochelobe_overflow}
In considering the dynamics of gas giants undergoing RLO, we make several assumptions, among which the most important are the following:

\begin{enumerate}
\item Gas escaping the planet likely forms a thin accretion disk around the star and transfers some or all of its angular momentum back to the planet before falling onto the star. We assume that the fraction of angular momentum lost from the orbit remains fixed. In reality, this fraction likely depends on a number of evolving properties of the system, including the planet's orbital period.

\item The planet has a circular orbit with a mean motion larger than the host star's rotational frequency and the total angular momentum of the planet-star system lies below the critical threshold for tidal equilibrium. As a result, the tide raised on the star lags behind the planet and tends to transfer angular momentum from the orbit to the stellar rotation, but tides cannot synchronize the star's rotation to the orbit.

\end{enumerate}

With these assumptions, we follow a simplified version of the derivation described in \cite{Rappaport1982} and \cite{2015ApJ...813..101V}. The orbital angular momentum $J$ is given approximately as 

\begin{equation}
J \approx M_{\rm p} \sqrt{G M_{\star} a},
\label{eqn:orbital_angular_momentum}
\end{equation}
where $M_{\rm p}$ is the planet's mass, $M_\star$ the star's mass, $G$ Newton's gravitational constant, and $a$ is the orbital semi-major axis. A change in angular momentum can arise from changes in $a$ or $M_{\rm p}$ (since $M_{\rm p} \ll M_{\star}$, we neglect changes to $M_{\star}$):
\begin{equation}
\frac{dJ}{dt} = \left(\frac{\partial J}{\partial t}\right)_{\rm tides} + \left(\frac{\partial J}{\partial t}\right)_{\rm \dot{M}_{\rm p}}.
\label{eqn:change_in_orbital_angular_momentum}
\end{equation}

We consider Darwin's model for tidal interaction and assume tidal dissipation can be modeled with a frequency-independent efficiency parameter for the star $Q_{\star}$ into which we have absorbed the stellar Love number \cite{Jackson2008Tidal}. There are many alternative formulations which involve more sophisticated assumptions about the nature of tidal dissipation and stellar internal structure (e.g., \cite{Essick2015Orbital}), but using a different tidal model would probably only modify the timescales for dynamical evolution and not substantially alter our results. 
\begin{equation}
\frac{1}{J}\left(\frac{\partial J}{\partial t}\right)_{\rm tides} = -\frac{9}{4} \left( \frac{G}{M_{\star}} \right)^{1/2} \frac{R_{\star}^5 M_{\rm p}}{Q_{\star}} a^{-13/2},
\label{eqn:tidal_decay}
\end{equation}
where $R_{\star}$ is the stellar radius.

Mass escaping the planet forms an accretion disk somewhat interior to the planet's orbit, at radius $\gamma^2 a$, giving the disk a specific orbital angular momentum $\gamma \sqrt{G M_{\rm \star} a}$. Some fraction $\delta$ of that disk mass may be lost to the system before transferring its angular momentum back to the planet's orbit, reducing the orbital angular momentum at a rate
\begin{equation}
\frac{1}{J}\left(\frac{\partial J}{\partial t}\right)_{\rm \dot{M}_{\rm p}} = \delta \gamma J \left( \frac{\dot{M_{\rm p}}}{M_{\rm p}} \right).
\label{eqn:loss_of_angular_momentum_from_system}
\end{equation}

The time derivative of $J$ can also be written as 
\begin{eqnarray}
\frac{1}{J}\frac{dJ}{dt} & \approx & \frac{1}{J} \left(\frac{\partial J}{\partial M_{\rm p}}\right)_{\rm M_{\rm \star}, a} \dot{M}_{\rm p} + \frac{1}{J} \left(\frac{\partial J}{\partial a}\right)_{\rm M_{\rm \star}, M_{\rm p}} \dot{a} \nonumber \\
& = & \left( \frac{\dot{M_{\rm p}}}{M_{\rm p}} \right) + \frac{1}{2} \left( \frac{\dot{a}}{a} \right),
\label{eqn:change_in_angular_momentum}
\end{eqnarray}
where we have ignored the effect of the change in the star's mass on $J$. Setting the two expressions for $\dot{J}/J$ equal and solving for $\dot{a}/a$ gives
\begin{equation}
\left(\frac{\dot{a}}{a}\right) = -2 \left(1 - \delta \gamma \right) \left( \frac{\dot{M}_{\rm p}}{M_{\rm p}}\right) + \frac{2}{J} \left( \frac{\partial J}{\partial t}\right)_{\rm tides}.
\label{eqn:initial_dadt}
\end{equation}

In our scenario, a hot Jupiter has its orbit decay according to Equation \ref{eqn:tidal_decay} until it comes into Roche-lobe contact, at which point 

\begin{equation}
R_{\rm p} = R_{\rm Roche},
\label{eqn:Rp_RRoche_equality}
\end{equation}
where $R_{\rm Roche}$ is given by 
\begin{equation}
R_{\rm Roche} = f \left( M_{\rm p}/M_{\star} \right)^{1/3} a
\label{eqn:RRoche}
\end{equation}
and $f\ (\sim 1)$ incorporates details of the planet's internal structure and constitution (cf. \cite{MurrayDermott1999}). 

Torques between the planet and disk, along with viscous dissipation within the disk, act in the opposite direction as the torque from the tide raised on the star \cite{1979MNRAS.186..799L}. Mass loss would choke off if the planet were no longer filling its Roche lobe, so the disk torque cannot push the planet outward of its Roche limit. At the same time, if the tide raised on the host star were to drive the planet inward through its Roche limit, the degree to which the planet overfilled its Roche lobe would increase, significantly increasing the rate of mass loss \cite{Ritter1988Turning}, the amount of mass in the accretion disk, and thus the strength of the outward torque. Consequently, in the case of stable RLO, a balance develops between the tidal and gas torques \cite{Priedhorsky1988Tidal}, and $R_{\rm Roche}$ is held very nearly equal to $R_{\rm p}$. 

We can work out how the planet's orbit evolves during RLO by differentiating Equation \ref{eqn:Rp_RRoche_equality} and incorporating Equations \ref{eqn:change_in_orbital_angular_momentum} and \ref{eqn:RRoche}:
\begin{eqnarray}
\xi \left( \frac{\dot{M}_{\rm p}}{M_{\rm p}} \right) + \left( \frac{\partial \ln R_{\rm p}}{\partial t} \right)_{\rm M_{\rm p}} & = & \left( \frac{\dot{a}}{a} \right) + \frac{1}{3} \left( \frac{\dot{M}_{\rm p}}{M_{\rm p}} \right) \nonumber \\ 
& = & -2 \left(1 - \delta \gamma \right) \left( \frac{\dot{M}_{\rm p}}{M_{\rm p}}\right) + \frac{2}{J} \left( \frac{\partial J}{\partial t}\right)_{\rm tides}.
\label{eqn:dRp_dt}
\end{eqnarray}
The first term on the left-hand side of Equation \ref{eqn:dRp_dt} incorporates the change in radius due to mass loss via RLO (with $\xi = \partial \ln R_{\rm p}/\partial \ln M_{\rm p}$), while the second term involves any other change in radius.  

Plugging in Equation \ref{eqn:tidal_decay} and re-arranging gives expressions for the mass loss rate and the accompanying evolution of the semi-major axis:
\begin{equation}
\left( \frac{\dot{M}_{\rm p}}{M_{\rm p}} \right) = -\eta^{-1} \left[ \frac{9}{2} \left( \frac{G}{M_{\star}} \right)^{1/2} \frac{R_{\star}^5 M_{\rm p}}{Q_{\star}} a^{-13/2} + \left( \frac{\partial \ln R_{\rm p}}{\partial t} \right)_{\rm M_{\rm p}} \right],
\label{eqn:RLO_mass_loss_rate}
\end{equation}
\begin{equation}
\left( \frac{\dot{a}}{a} \right) = \eta^{-1} \left[\left( \frac{1}{3} - \xi \right) \frac{9}{4} \left( \frac{G}{M_{\star}} \right)^{1/2} \frac{R_{\star}^5 M_{\rm p}}{Q_{\star}} a^{-13/2} + \left( 1 - \delta \gamma \right) \left( \frac{\partial \ln R_{\rm p}}{\partial t} \right)_{\rm M_{\rm p}} \right],
\label{eqn:a_evolution_rate}
\end{equation}
where $\eta \equiv \xi/2 + 5/6 - \delta \gamma$. 
Solving these equations together requires a numerical model that can account for the planetary and stellar evolution, which we implement using MESA below. 

Of course, this derivation is predicated on the assumption that the mass loss remains stable, which requires $\eta > 0$. Otherwise, $\dot{M}_{\rm p}$ would pass through unphysically large (negative) values or even reverse sign. This requirement translates into
\begin{equation}
\delta \gamma < \xi/2 + 5/6.
\label{eqn:stability_condition}
\end{equation}
This equation puts an upper limit on the fraction of angular momentum lost from the system during mass transfer and agrees with the result discussed in \cite{2015ApJ...813..101V}. For instance, consider $\gamma \approx 1$ (i.e., the accretion disk orbits very near the planet) and $\xi \approx 0$ (i.e. $R_{\rm p}$ is insensitive to $M_{\rm p}$ as we discuss next). In that case, no more than 5/6 (= $0.8\bar{3}$) of the mass escaping from the planet can be lost from the system without returning its angular momentum. Otherwise, $R_{\rm Roche}$ would decrease more slowly than $R_{\rm p}$, and mass loss would become unstable. For the suite of simulations they explored, \cite{2015ApJ...813..101V} found that $\delta \gamma$ larger than 0.6 to 0.8 (depending on the planet modeled) could result in unstable mass transfer, consistent with our discussion here. 

Even in the case that no mass is lost from the system, we would still expect some loss of orbital angular momentum since gas accreted by the host star carries a specific angular momentum $\sqrt{G M_\star R_\star}$. For a planet undergoing RLO at semi-major axis $a$, this loss of orbital angular momentum amounts to $\gamma = \sqrt{R_\star/a}$. For a hot Jupiter orbiting a Sun-like star and just encountering its Roche limit for the first time, $a = a_{\rm Roche} \approx 0.01\ {\rm AU} \approx 2\ R_\star$, and $\gamma = \sqrt{1/2}$. In a case like this one, $\delta \gamma$ is not a constant during RLO (as assumed in this study) but evolves as the orbit evolves. We leave exploration of the influence of an evolving $\delta \gamma$ for future work, but we do consider a $\delta \gamma = 0.5$ below.

Without solving the equations in detail, we can develop an intuition for how a overflowing hot Jupiter should evolve by applying several simplifications. First, assume the $\partial R_{\rm p}/\partial t$ in both equations are negligible -- \cite{Arras2006Thermal} calculated that, at ages $\geq$ 1 Gyrs, hot Jupiters take Gyrs more for their radii to contract by $\sim$ 10\%. Upon encountering its Roche limit, a hot Jupiter will begin losing mass to the star, and whether the orbit expands ($da/dt > 0$) or decays ($da/dt < 0$) depends on how $\left( \frac{1}{3} - \xi \right)$ evolves. In other words, the evolution of the orbit depends on the evolution of the planet's density, as we might expect. For hot Jupiters, in fact, $\xi \sim 0$ while $M_{\rm p} \sim 1\ {\rm M_{Jup}}$ \cite{Fortney2007Planetary}. 

With these approximations, upon encountering its Roche limit, a hot Jupiter will begin losing mass to the star, and its density will drop, which causes its Roche limit to move outward. Consequently, the torques in the accretion disk should drive the planet outward, which will follow the Roche limit very closely. At larger $P$ (and/or smaller $M_{\rm p}$), the influence of tides raised on the star will decline, which will reduce the mass transfer rate. 

As the hot Jupiter loses its atmosphere, eventually its core (if it has one) begins to dominate its mass, and the density increases as mass is lost and the Roche limit can retreat inward. If the tidal torque is sufficiently strong at that point, the remaining planet can follow the Roche limit back in, and mass loss will continue until the planet encounters the stellar surface, at which point, accretion of the planet may produce very bright optical and X-ray transient signals \cite{2012MNRAS.425.2778M}. 

If the initial orbital expansion drives the planet far enough out, the tidal torque may become small enough that mass loss rate tapers off, as shown in Equation \ref{eqn:RLO_mass_loss_rate}, which would presumably leave the planet with a gaseous envelope. Radiative cooling dominates the time evolution of $R_{\rm p}$ and gives $\partial \ln R_{\rm p}/\partial t < 0$, meaning that term only reduces the mass loss rate.

However, RLO is not the only process that removes planetary mass -- short-period gas giants are also prone to photoevaporative mass loss, in which heating of a planet's upper atmosphere by X-ray and ultraviolet (XUV) drives a hydrodynamic outflow  \cite{VidalMadjar2003Extended,2014ApJ...783...54K}. Several studies, including \cite{Lopez2013Role}, have shown that this evaporative mass loss can completely remove the atmosphere from a Neptune/sub-Neptune-like planet but has little effect on the total mass of a hot Jupiter. 

\cite{Erkaev2007Roche} provided a useful parameterization for energy-limited evaporative mass loss $\dot{M}_{\rm p, evap}$ incorporating the effect of tides:
\begin{equation}
\dot{M}_{\rm p, evap} = -\frac{\epsilon \pi R_{\rm p}^3 F_{\rm XUV}}{G M_{\rm p} K_{\rm tide}},
\label{eqn:evaporative_mass_loss}
\end{equation}
where $F_{\rm XUV}$ is the stellar XUV flux at the distance of the planet, which tends to fall off for Sun-like stars as the stars age \cite{Ribas2005Evolution}. $R_{\rm p}$ is the planetary radius, and $\epsilon$ is the fraction of incoming XUV energy powering atmospheric escape, typically 10\% \cite{Owen2012Planetary}. Since gas outflowing from a planet close to its host star only has to reach the Roche lobe to escape, the required escape energy is smaller than if the gas had to escape to infinity, and $K_{\rm tide}$ represents this reduction in the escape energy, ranging from zero (a planet filling its Roche lobe) up to unity (a planet very far from filling its Roche lobe). 

Among other studies, \cite{2009ApJ...693...23M} employed a detailed radiative and hydrodynamic model to show that Equation \ref{eqn:evaporative_mass_loss} provides a reasonable approximation for the mass loss rate. However, that model (and Equation \ref{eqn:evaporative_mass_loss}) are predicated upon the assumption that the atmosphere at the XUV photosphere is still gravitationally bound to the planet. Instead, in the case that a planet fills its Roche lobe, the photosphere is not bound by definition, and the atmosphere structure adjusts as mass loss sets in. In fact, $K_{\rm tide}$ in Equation \ref{eqn:evaporative_mass_loss} decreases without limit as $R_{\rm p} \rightarrow R_{\rm Roche}$, and so the predicted $\dot{M}_{\rm p, evap}$ blows up. Thus, the nature of photoevaporative mass loss is not clear for a planet in RLO. 

Indeed, RLO and evaporative mass loss are not two distinct processes occurring in isolation from one another; rather, they are endmembers along a spectrum of atmospheric escape. The former operates when a planet fills its Roche lobe, while the latter can operate even when the planet is far from filling the Roche lobe (e.g., for Pluto in our solar system -- \cite{Watson1981Dynamics}). However, as pointed out in \cite{Ritter1988Turning} and \cite{Li2010WASP12b}, since planetary (and stellar) atmospheres do not terminate at a hard boundary but gradually taper off into space, RLO can operate even if the base of a planet's atmosphere is a few atmospheric scale heights from the Roche lobe. The transition between RLO and evaporative mass loss probably plays an important in sculpting the atmospheres of the remnants of hot Jupiters and will be explored in future work. \cite{2015ApJ...813..101V} did include photoevaporation in their models using Equation \ref{eqn:evaporative_mass_loss}, but it is not clear how that study circumvented the singularity during RLO. In the absence of a photoevaporative mass loss model that clearly applies during RLO, we do not include evaporative mass loss in our calculations.

Turning back to RLO and tidal interactions, all of this planetary evolution is taking place in the presence of an evolving star, and, as we show below, the rate of stellar evolution may exceed those of orbital evolution and mass transfer. In this case, we expect that, once RLO begins, the planet will continue losing mass until the star leaves the main sequence, at which point a Sun-like star will enter the post-main sequence and may accrete the planet during its bloated red giant phase \cite{2009ApJ...700..832C}.

Thus, we expect four distinct outcomes for close-in planets involving tidal decay and RLO:
\begin{enumerate}
\item \underline{Little to No Tidal Decay}: Tidal decay occurs slowly enough that the close-in planet does not encounter its Roche limit during the main sequence lifetime of the host star. Very likely such planets are accreted during the post-main sequence, but observational evidence for this accretion among red giant stars, including unusually high spin rates and Li content, is currently ambiguous \cite{2012ApJ...757..109C}.

\item \underline{Complete Accretion of the Planet}: The planet does encounter its Roche limit and undergoes mass transfer, but mass loss and the subsequent reduction in planetary density do not move the Roche limit out very far, leaving the tidal torque and, consequently, the mass loss at relatively large values. Thus, the planet may quickly lose its gaseous envelope, and the remaining core will spiral into the star, all during the main sequence. As discussed above, the large rotation rates for \kepler\ stars not seen to host close-in planets \cite{2013ApJ...775L..11M} may provide evidence for this scenario.

\item \underline{Incomplete RLO of the Planet's Atmosphere}: Mass transfer begins and moves the planet far enough out that it is not completely accreted by the star during the main-sequence. However, loss of the planet's atmosphere and contraction of its radius occur slowly, and the planet never stops shedding mass either through RLO or photoevaporation during the main sequence lifetime of the star. Planets evolving in this way should be found near their current Roche limits, and the fact that such a population is not observed might argue this scenario occurs rarely or not at all.

\item \underline{Complete RLO of the Atmosphere Alone}: This scenario is similar to the third scenario above and distinct from the first scenario in that accretion of the remnant planet does not occur during the main sequence. During mass transfer, the planet's density does turn over, stranding the remnant at a maximum Roche limit period during its evolution, $P_{\rm Roche, max}$. As we show in the next section, this maximum orbital period depends sensitively on the core mass, which means that the period of a planet that is the remnant of a hot Jupiter, whether gas-rich or entirely solid, will be largely determined by the mass of its solid core. 

\end{enumerate}

It is not clear that gas giants always have solid cores, and recent work has suggested that gas giants may even begin with solid cores but lose them through dissolution of the core in the gaseous envelope \cite{2012ApJ...745...54W}. Any of the above scenarios could still apply to the case of a coreless gas giant, but, of course, complete removal of the atmosphere means removal of the planet altogether. The mass-radius relation of a coreless gas giant would likely produce different orbital evolution than discussed here, but we do not explore that case in what follows. 

Which of these scenarios applies depends in fairly simple ways on the combination of a planet's initial mass and period, its core mass, $Q_\star$, and the product $\delta \gamma$. The suite of simulations described in the next section illustrates examples of each.

\section{RLO Results from MESA}
\label{sec:RLO_Results_from_MESA}
\begin{figure}
\includegraphics[width=\textwidth]{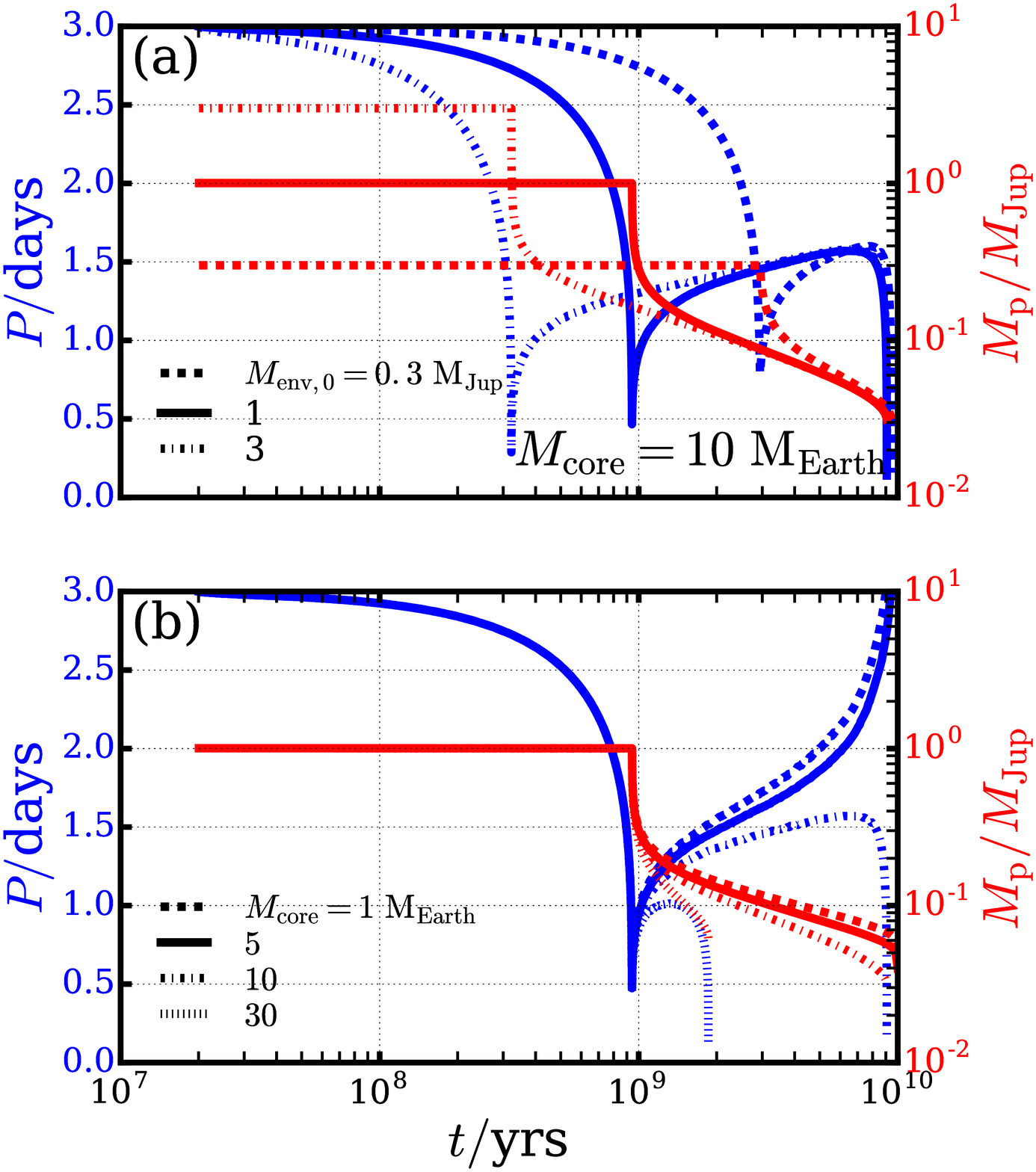}
\caption{Mass (red lines) and orbital (blue lines) evolution for hot Jupiter systems with initial periods $P_{0} =$ 3 days and $Q_\star = 10^5$ amd a variety of initial envelope masses $M_{\rm env,\ 0}$ and core masses $M_{\rm core}$. The different linestyles indicate different planetary parameters. (a) Hot Jupiters with $M_{\rm env,\ 0} =$ 0.3, 1, and 3 ${\rm M_{Jup}}$ and $M_{\rm core}$ fixed at 10 ${\rm M_{Earth}}$. (b) Hot Jupiters with $M_{\rm env,\ 0} = 1\ {\rm M_{Jup}}$ and $M_{\rm core} = $ 1, 5, 10, and 30 ${\rm M_{Earth}}$. These calculations assume $\delta \gamma = 0$, i.e. the orbital angular momentum is completely conserved.}
\label{fig:plot_mass_orbital_evolution_variable-Mp0-Mcore}
\end{figure}

Figure \ref{fig:plot_mass_orbital_evolution_variable-Mp0-Mcore} (a) illustrates the evolution for planets with a range of initial envelope masses $M_{\rm env, 0} = \left[0.3, 1, 3\right]$ ${\rm M_{Jup}}$ but all with the same core $M_{\rm core} = 10\ {\rm M_{Earth}}$. In (b), the planets all begin with the same $M_{\rm env, 0} = 1\ {\rm M_{Jup}}$ but with $M_{\rm core} = \left[1, 5, 10, 30\right]\ {\rm M_{Earth}}$. All host stars have $M_\star = 1\ {\rm M_{Sun}}$, a Sun-like initial rotation velocity (2 km/s), and $Q_\star = 10^5$. All systems begin with initial periods $P_0 = 3\ {\rm days}$ and evolve from 20 Myrs to 10 Gyrs after formation. We assume solar metallicity for the star and the planet's atmosphere. Irradiation from the star is deposited within the planet's atmosphere at a fixed column depth, 100 g/cm$^2$, as in \cite{2015ApJ...813..101V}. We use the ``implicit'' RLO mass loss scheme in MESA. The host star sheds angular momentum through a rotationally-enhanced wind, as described in \cite{2013ApJS..208....4P}, and we hold the stellar radius fixed throughout the evolution. A consequence of this latter assumption is that we may underestimate the tidal torque, especially as the star leaves the main sequence and expands. For Figure \ref{fig:plot_mass_orbital_evolution_variable-Mp0-Mcore}, we also assume completely conservative mass transfer, i.e. $\delta \gamma = 0$. Additional model details are given in our MESA ``inlists'' and results files, all available at http://www.astrojack.com/research.

Although the planets in (a) begin with a range of initial masses and encounter the Roche limit at different times, the evolutionary curves all converge at late times because $\dot{a}$ and $\dot{M_{\rm p}}$ both depend on $\left( \partial J/\partial t \right)_{\rm tides}$, which scales with $M_{\rm p}$. Thus, the planets with larger $M_{\rm env, 0}$ experience more rapid mass loss and orbit evolution, which eventually slow down as $M_{\rm p}$ drops, allowing the planets with smaller initial masses to catch up. 

Moreover, by late times ($t \gtrsim 5\ {\rm Gyrs}$), the planets have lost sufficient mass that they have evolved from hot Jupiters, with envelope mass fractions $f_{\rm env} = M_{\rm env}/M_{\rm p} \sim 1$, to super-Earths/sub-Neptunes, with $f_{\rm env} \le 0.5$. \cite{Lopez2014Understanding} showed that the radii of such planets are almost entirely determined by the value of $f_{\rm env}$. Mass transfer of a planet's gaseous envelope reduces $f_{\rm env}$, so, as the $f_{\rm env}$-values converge, the planets' densities and Roche limits converge. 

Figure \ref{fig:plot_mass_orbital_evolution_variable-Mp0-Mcore} (b) drives home the key role played by $M_{\rm core}$ in determining the evolution. For the planets with $M_{\rm core} \le 10\ {\rm M_{Earth}}$, mass transfer reduces the planetary densities to small values, driving the Roche limit and the planets back out to $P = 3$ days before the simulations end. In panel (b), those planets with $M_{\rm core} \le 10\ {\rm M_{Earth}}$ do not shed their entire envelopes during the course of the simulation because the orbital expansion that accompanies the mass transfer significantly reduces $\left( \partial J/\partial t \right)_{\rm tides}$. 

\begin{figure}
\includegraphics[width=\textwidth]{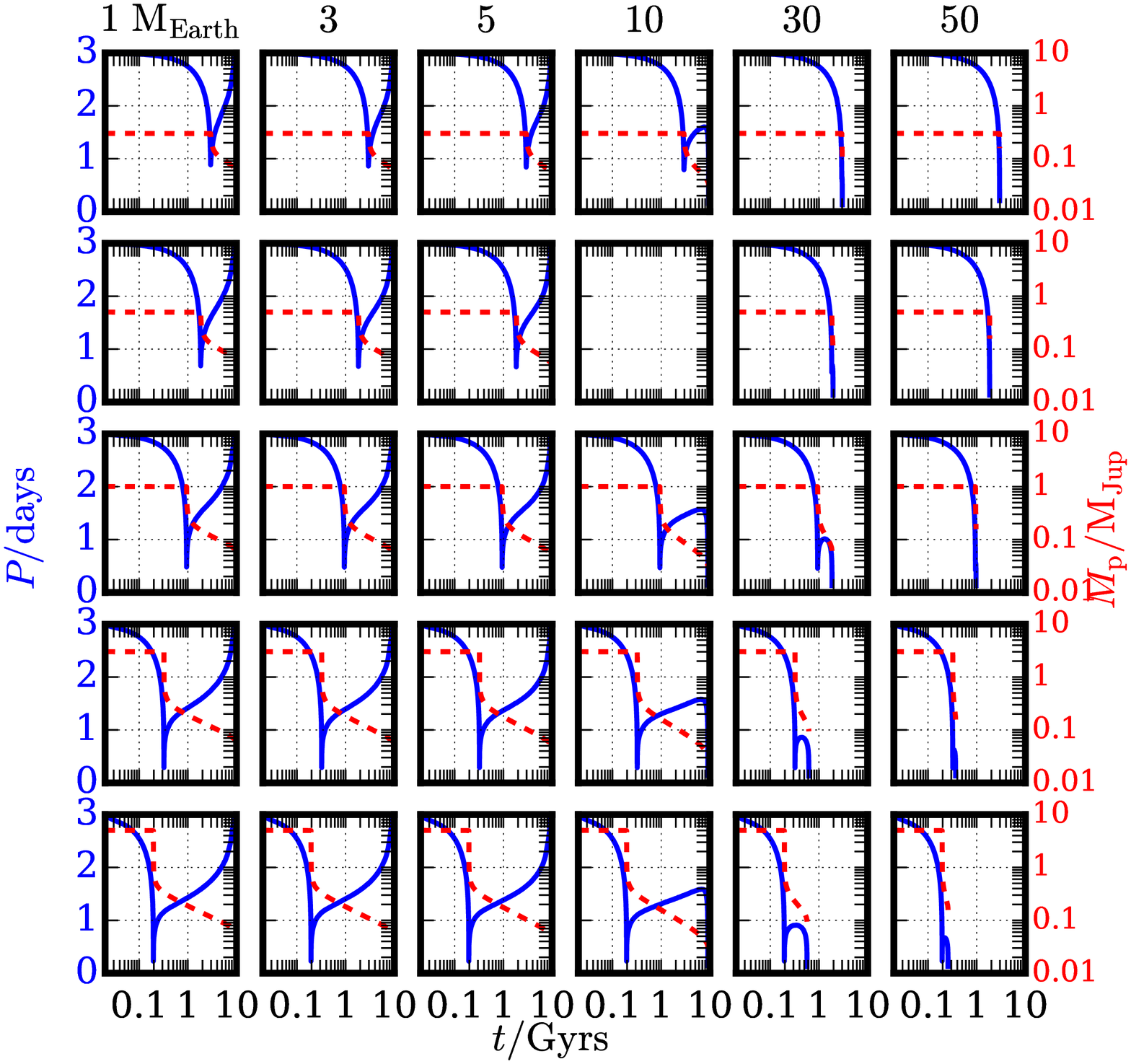}
\caption{Tidal evolution and mass transfer of planets with (top to bottom) $M_{\rm env, 0} =$ 0.3, 0.5, 1, 3, and 5 ${\rm M_{Jup}}$ and (left to right) $M_{\rm core} =$ 1, 3, 5, 10, 30, and 50 ${\rm M_{Earth}}$. All host stars have masses $M_\star = 1\ {\rm M_{Sun}}$ and $Q_\star = 10^5$. The combination $M_{\rm env, 0} = 0.5\ {\rm M_{Jup}}$/$M_{\rm core} = 10\ {\rm M_{Earth}}$ is missing due to convergence issues.}
\label{fig:plot_mass_orbit_evolution_grid_qs5}
\end{figure}

Figure \ref{fig:plot_mass_orbit_evolution_grid_qs5} demonstrates a wide range of evolution for a grid of $M_{\rm env, 0}$ and $M_{\rm core}$. (We encountered insoluable convergence problems for the combination $M_{\rm env, 0} = 0.5\ {\rm M_{Jup}}$/$M_{\rm core} = 10\ {\rm M_{Earth}}$.) As shown Figure \ref{fig:plot_mass_orbital_evolution_variable-Mp0-Mcore}, the key parameter determining the degree of orbital expansion is $M_{\rm core}$. For example, all planets with $M_{\rm core} = 10\ {\rm M_{Earth}}$ back out to $P = 1.4$ days (before tumbling into their host stars). This result agrees with those of \cite{2015ApJ...813..101V} for the same core mass. Planets with less massive cores continuing moving back out toward $P = 3$ days and do not shed their entire atmospheres during the 10-Gyr simulation. By contrast, planets with $M_{\rm core} \ge 30\ {\rm M_{Earth}}$ back out only a little or not at all with onset of mass transfer. 

\begin{figure}
\includegraphics[width=\textwidth]{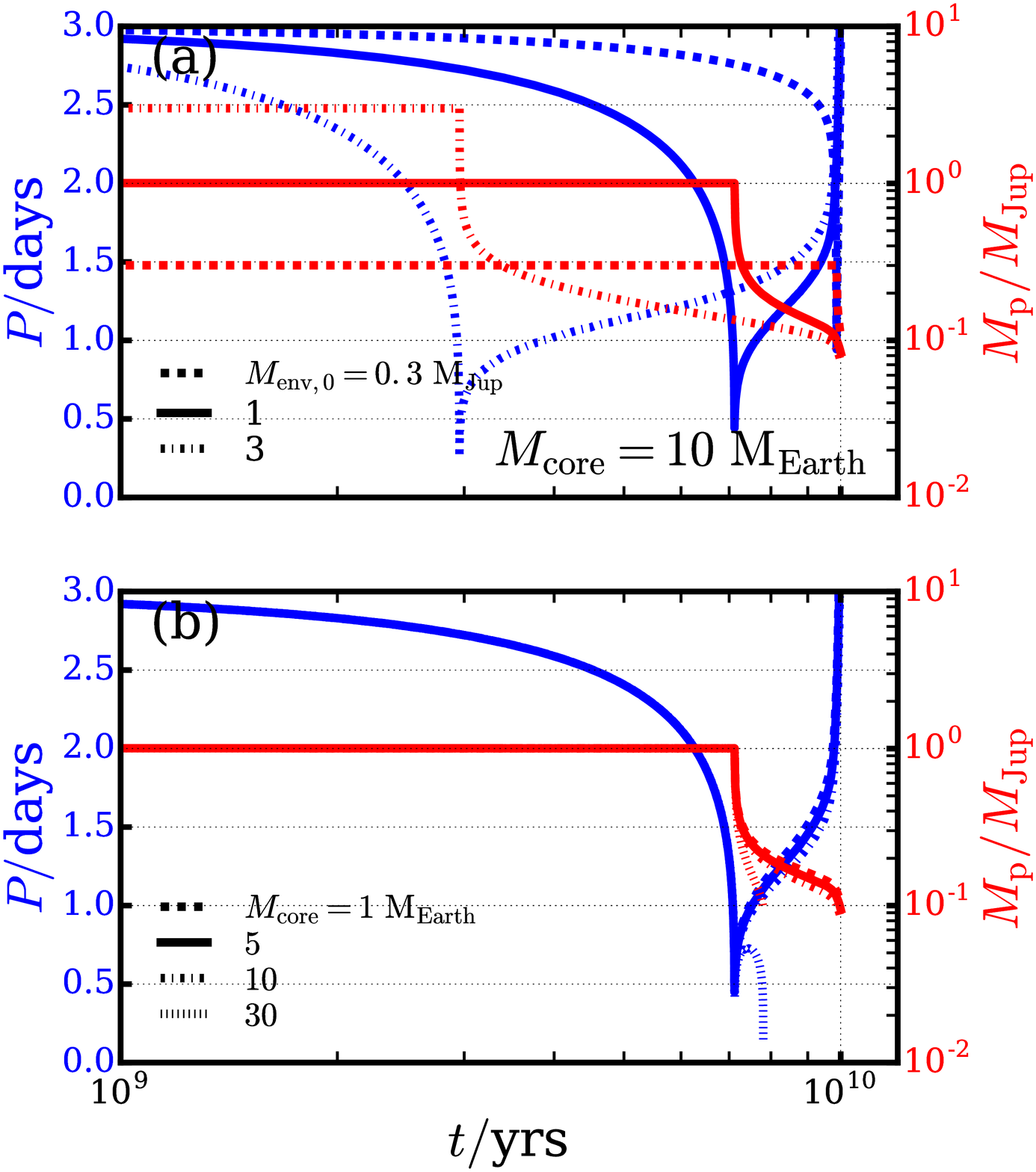}
\caption{Similar to Figure \ref{fig:plot_mass_orbital_evolution_variable-Mp0-Mcore}, except that $Q_\star = 10^6$. The x-axis also spans a different range.}
\label{fig:plot_mass_orbital_evolution_variable-Mp0-Mcore_qs6}
\end{figure}

Since it helps determine the rate of angular momentum transfer, $Q_\star$ also factors into the fate of the planet, and Figure \ref{fig:plot_mass_orbital_evolution_variable-Mp0-Mcore_qs6} shows evolution curves similar to those in Figure \ref{fig:plot_mass_orbital_evolution_variable-Mp0-Mcore} but for $Q_\star = 10^6$. In this case, none of the modeled planets in either panel (a) or (b) encounters their Roche limits until much later than in Figure \ref{fig:plot_mass_orbital_evolution_variable-Mp0-Mcore}, and so by the end of the simulation, the planets still retain substantial gaseous envelopes. The departure of the star from the main sequence increases its luminosity, which strongly heats the planets' remaining atmospheres, driving the planets' bulk densities to very small values and the Roche limits out. However, even in this case, the evolution for all planets with a given core mass converge. 

The upshot of these simulations is that a planet's core mass can play a dominant role in setting the orbital evolution of an overflowing planet. As long as mass transfer occurs quickly enough that stellar evolution is not a factor, we expect that a planetary mass-radius relationship that incorporates the dependence on core mass should allow us to predict at least the period at which orbital expansion of an overflowing planet halts. If tidal decay ceased at that point, such a mass-radius relation would indicate exactly where we could find the remnant of a gaseous planet, given the remnant's mass (and density).

Unfortunately, comparing Figures \ref{fig:plot_mass_orbital_evolution_variable-Mp0-Mcore} and \ref{fig:plot_mass_orbital_evolution_variable-Mp0-Mcore_qs6}, it seems that, for conservative mass transfer, transfer rates large enough to nearly completely remove an atmosphere require tidal decay rates that produce rapid orbital decay of the remnant, destroying the evidence (the ``Complete RLO and Accretion'' scenario above). 

Whether such a scenario commonly occurs is unclear since the appropriate values for $Q_\star$ are unknown, and it is not clear the mass transfer is completely conservative. As discussed above, it also seems plausible that some form of photoevaporative atmospheric escape should contribute to the mass loss, although it is not obvious what form it should take. Altogether, these uncertainties mean that the timescales for mass loss and tidal decay are not clear. 

\begin{figure}
\includegraphics[width=\textwidth]{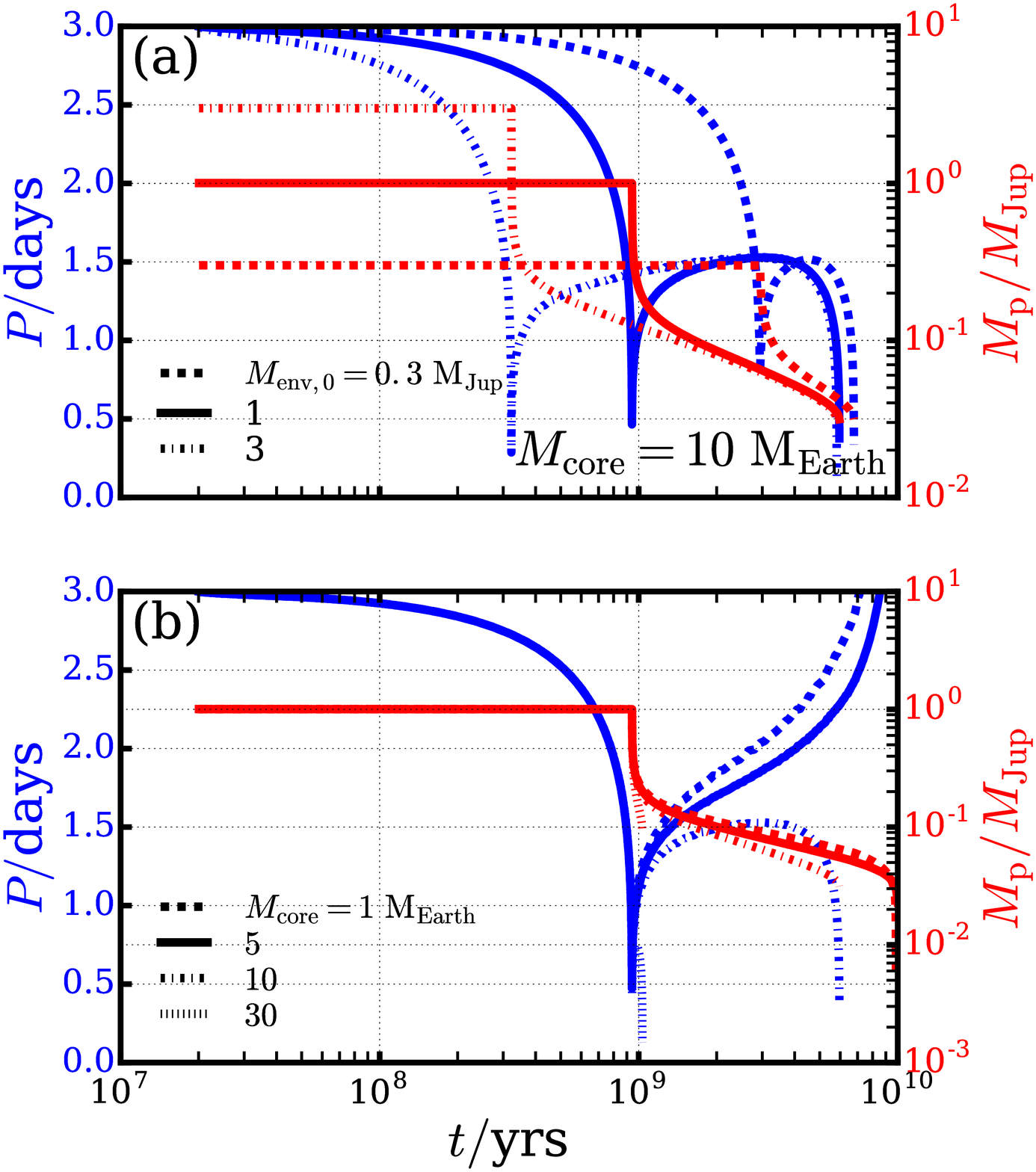}
\caption{Mass (red lines) and orbital (blue lines) evolution for hot Jupiter systems with initial periods $P_{0} =$ 3 days and $Q_\star = 10^5$ amd a variety of initial envelope masses $M_{\rm env,\ 0}$ and core masses $M_{\rm core}$. The different linestyles indicate different planetary parameters. (a) Hot Jupiters with $M_{\rm env,\ 0} =$ 0.3, 1, and 3 ${\rm M_{Jup}}$ and $M_{\rm core}$ fixed at 10 ${\rm M_{Earth}}$. (b) Hot Jupiters with $M_{\rm env,\ 0} = 1\ {\rm M_{Jup}}$ and $M_{\rm core} = $ 1, 5, 10, and 30 ${\rm M_{Earth}}$. These calculations assume $\delta \gamma = 0.5$, i.e half the orbital angular momentum carried by the escaping gas is lost.}
\label{fig:plot_mass_orbital_evolution_variable-Mp0-Mcore_qs5_del1-gam0p5}
\end{figure}

Allowing for some loss of orbital angular momentum does not qualitatively modify these results. Instead, such loss simply modifies the timescales for the evolution to take place. Figure \ref{fig:plot_mass_orbital_evolution_variable-Mp0-Mcore_qs5_del1-gam0p5} illustrates cases similar to Figure \ref{fig:plot_mass_orbital_evolution_variable-Mp0-Mcore} except that we take $\delta \gamma = 0.5$, i.e. half the orbital angular momentum carried by the escaping gas is lost, and $\eta^{-1}$ is larger in Equations \ref{eqn:RLO_mass_loss_rate} and \ref{eqn:a_evolution_rate}.

Figure \ref{fig:plot_mass_orbital_evolution_variable-Mp0-Mcore_qs5_del1-gam0p5} closely resembles Figure \ref{fig:plot_mass_orbital_evolution_variable-Mp0-Mcore}, except that the mass loss and orbital evolution proceed more quickly. For example, in panel (a), the planet with $M_{\rm env, 0} = 1\ M_{\rm Jup}$ reaches $P_{\rm Roche, max} = 1.5$ days by $t = 4$ Gyrs, instead of 9 Gyrs as in Figure \ref{fig:plot_mass_orbital_evolution_variable-Mp0-Mcore} (a). The fact that the orbital expansion shown in Figure \ref{fig:plot_mass_orbital_evolution_variable-Mp0-Mcore} does not reverse until nearly the end of the star's main sequence lifetime (when the star is brighter) also means that the planets' atmospheres are hotter and therefore more distended than those shown in Figure \ref{fig:plot_mass_orbital_evolution_variable-Mp0-Mcore_qs5_del1-gam0p5} (a). Consequently, $P_{\rm Roche, max}$ is slightly larger. Likewise, the $Q_\star = 10^6$/$\delta \gamma = 0.5$ case (not shown) closely resembles the $Q_\star = 10^6$/$\delta \gamma = 0$ (Figure 6): once it starts, RLO proceeds more rapidly for the former than for the latter case, but, qualitatively, the evolution for both is very similar. These results suggest that, as long as the mass transfer is stable, non-conservation of angular momentum does not qualitatively modify the simple scenario outlined above. 

Thus, assuming there is some remnant left behind after removal of a hot Jupiter's atmosphere, the mass loss and angular momentum transfer timescales are of secondary importance in determining the properties and orbit of the remnant. Instead, we suggest that the mass-radius relationship dominates, and we expect to find remnants stranded near or interior to $P_{\rm Roche, max}$, the point at which the tidal interaction becomes weakest. 

A key feature of this hypthothesis is that, in order to denude a gas planet's solid core, loss of the atmosphere must continue, even as the Roche limit retreats inward of the planet's orbit. Since, in principle, this retreat of the Roche limit means the planet is no longer in Roche-lobe contact, RLO must taper off. However, at that point, we suggest photoevaporation can take over and remove much of the remaining atmosphere to produce a planet that not currently observed in or near RLO. Otherwise, a population of planets with $P = P_{\rm Roche}$ should be observed, and it is not -- Figure \ref{fig:P-PRoche}.

Of course, our simulations here indicate more complicated evolution can occur, but in the interest of providing a clear, testable prediction, we next consider how the Roche limit evolves as the planet loses its gaseous envelope and whether there is evidence for a population of low-mass, short-period planets near periods where we would expect remnants.

\section{Evolution of the Roche Limit for an Overflowing Gaseous Planet}
\label{sec:Evolution_of_the_Roche_Limit_for_a_Disrupting_Gaseous_Planet}
Applying a full planetary evolution model, \cite{Lopez2014Understanding} studied the dependence of the radii of sub-Neptunes/super-Earths (with $M_{\rm p} \le 20\ {\rm M_{Earth}}$) on the planets' envelope and core masses, stellar insolation, and age. They fit the following series of power-laws to provide an analytic mass-radius relation:
\begin{equation}
R_{\rm p} \approx 2.06\ {\rm R_{Earth}}\ \left( \dfrac{M_{\rm p}}{\rm M_{Earth}} \right)^{-0.21} \left( \dfrac{f_{\rm env}}{5\%} \right)^{0.59} \left( \dfrac{F_{\rm p}}{\rm F_{Earth}} \right)^{0.044} \left( \dfrac{\rm age}{\rm 5\ Gyrs} \right)^{-0.18} + R_{\rm core}\  ,
\label{eqn:LopezFortney2014_subNeptune_relation}
\end{equation}
where $F_{\rm p}$ is the stellar insolation received by the planet. Equation \ref{eqn:LopezFortney2014_subNeptune_relation} involves a number of approximations, including neglecting the contribution to $R_{\rm p}$ of a radiative outer atmosphere, which is usually small ($0.1\ {\rm R_{Earth}}$). The last term represents the radius of the solid core, which is insensitive to the exact proportion of iron and rock and is given in \cite{Lopez2014Understanding}:
\begin{equation}
R_{\rm core} \approx \left( \dfrac{M_{\rm core}}{\rm M_{Earth}} \right)^{0.25}\ {\rm R_{Earth}}. 
\label{eqn:LopezFortney2014_rockyplanet_relation}
\end{equation}

\begin{figure}
\includegraphics[width=\textwidth]{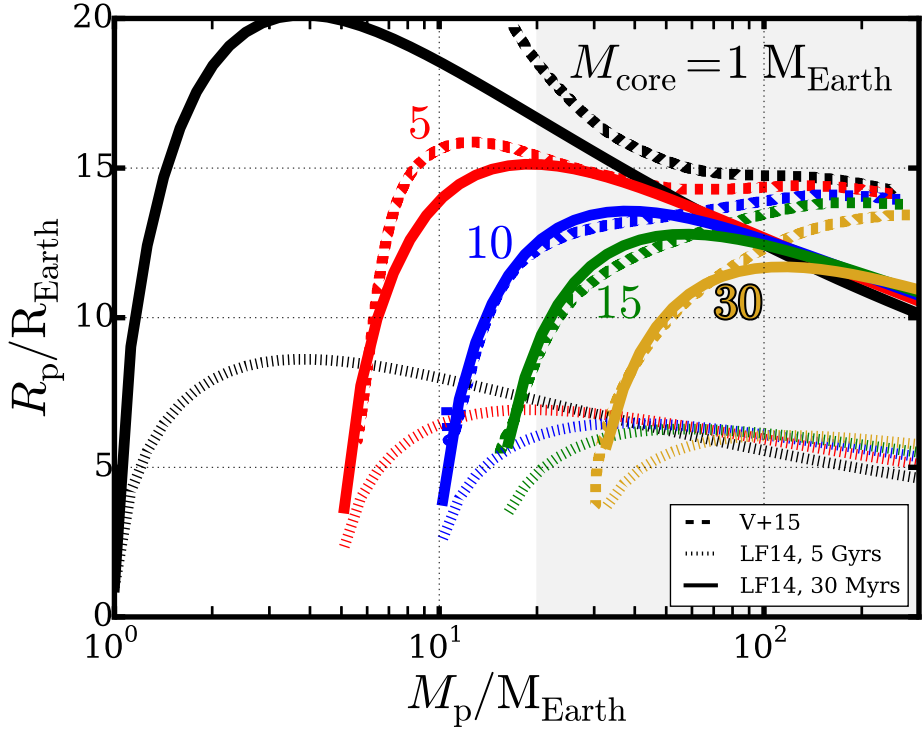}
\caption{Mass-radius relations from \cite{2015ApJ...813..101V}, labeled ``V+15'', and from \cite{Lopez2014Understanding} and Equation \ref{eqn:LopezFortney2014_subNeptune_relation}, labeled ``LF14''. For the LF14 relations, we have assumed planetary ages of 5 Gyrs (dotted lines) and 30 Myrs (solid lines). The color of a line corresponds to an assumed $M_{\rm core}$ (as labeled), and the gray region shows where we have extrapolated outside the range of $M_{\rm p}$ considered in \cite{Lopez2014Understanding}.}
\label{fig:compare_mass-radius_relations}
\end{figure}

Figure \ref{fig:compare_mass-radius_relations} compares the mass-radius relationship given by Equation \ref{eqn:LopezFortney2014_subNeptune_relation} to that given in Figure 7 of \cite{2015ApJ...813..101V} based on the RLO calculations for that paper. (It is important to note that the power-law fits in \cite{Lopez2014Understanding} were made only for $M_{\rm p} \le 20\ {\rm M_{Earth}}$, and the gray region in the figure shows where we have extrapolated beyond that point.) 

As indicated in \cite{2014ApJ...793L...3V}, the age of a planet becomes less important in determining the radius as the planet ages, and that study considers a fiducial age of 5 Gyrs, at which point the radius is insensitive to the initial entropy assumed for the planet's interior. Interestingly, the radii for this fiducial age lie considerably below the radii given in \cite{2015ApJ...813..101V} for planets in RLO. 

In fact, the reason for this disagreement is addressed in \cite{2014ApJ...793L...3V}: all other things being equal, larger planets cool more slowly, owing to their larger volume to surface area ratio, and it seems that planets arriving at a given mass via RLO retain higher internal entropies than planets born with that mass. Indeed, the entropies estimated by MESA for the convective interiors of our modeled planets from the previous section at a given $M_{\rm p}$ and age do exceed the corresponding entropies shown in Figure 4 of \cite{2014ApJ...793L...3V}. 

By using a fiducial age of 30 Myrs in Equation \ref{eqn:LopezFortney2014_subNeptune_relation} instead (solid lines in Figure \ref{fig:compare_mass-radius_relations}), we can bring the radius estimates into reasonable agreement with those of \cite{2015ApJ...813..101V}, at least at the lowest masses, which is the most important portion of the curve for determining $P_{\rm Roche, max}$. (Other similar, fiducial ages give worse qualitative agreement between the curves.)

We can re-write Equation \ref{eqn:LopezFortney2014_subNeptune_relation} using this new fiducial age of 30 Myrs. Since $f_{\rm env}$ drops while $M_{\rm core}$ remains constant during RLO, with $M_{\rm p} = M_{\rm core}/\left(1 - f_{\rm env}\right)$, it is worth re-casting the equation in terms of those variables:
\begin{equation}
R_{\rm p} \approx 5.17\ {\rm R_{Earth}} \left( \dfrac{M_{\rm core}}{M_{\rm Earth}} \right)^{-0.21} \left( \dfrac{f_{\rm env}}{0.05} \right)^{0.59} \left( 1 - f_{\rm env} \right)^{0.21} + \left( \dfrac{M_{\rm core}}{M_{\rm Earth}} \right)^{0.25}\ {\rm R_{Earth}}.
\label{eqn:recast_LopezFortney2014_subNeptune_relation}
\end{equation}
Here we have assumed $F_{\rm p} = F_{\rm Earth}$ since the insolation has only a small effect on the radius.

\begin{figure}
\includegraphics[width=\textwidth]{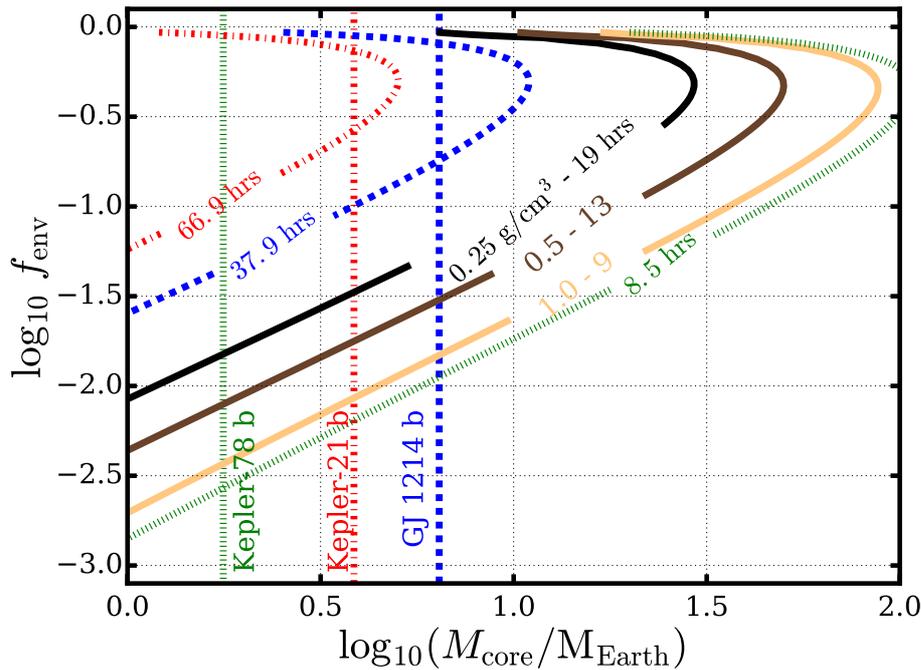}
\caption{The copper-to-black, solid contours show density and $P_{\rm Roche}$ based on the mass-radius relationship Equation \ref{eqn:LopezFortney2014_subNeptune_relation} from \cite{Lopez2014Understanding}. The leftmost number labeling each contour is the density in g/cm$^3$ and the rightmost the corresponding $P_{\rm Roche}$ in hours. The vertical blue, red, and green dashed lines show the core masses estimated using Equation \ref{eqn:recast_LopezFortney2014_subNeptune_relation} for GJ 1214 b, Kepler-21 b, and Kepler-78 b, respectively. The contour corresponding in color and linestyle shows the $P_{\rm Roche} = P$ curve for that planet's observed orbital period $P$ but not the planet's density.}
\label{fig:LopezFortney2014_density-PRoche_contours}
\end{figure}

With this mass-radius relation in hand, we can solve for planetary density and therefrom for $P_{\rm Roche}$ as functions of $M_{\rm core}$ and $f_{\rm env}$. As in \cite{Lopez2014Understanding}, we can estimate $f_{\rm env}$ by assuming $M_{\rm p} \approx M_{\rm core}$ and, for planets with small $f_{\rm env}$, we can compare their current orbital periods to the $P_{\rm Roche, max}$ expected for a given $M_{\rm core}$-value. Figure \ref{fig:LopezFortney2014_density-PRoche_contours} shows contours of $P_{\rm Roche}$ as a function of $M_{\rm core}$ and $f_{\rm env}$ and the $M_{\rm core}$-values for three example, short-period planets, GJ 1214 b \cite{2009Natur.462..891C}, Kepler-21 b \cite{2012ApJ...746..123H}, and Kepler-78 b \cite{2013ApJ...774...54S}. 

\begin{figure}
\includegraphics[width=\textwidth]{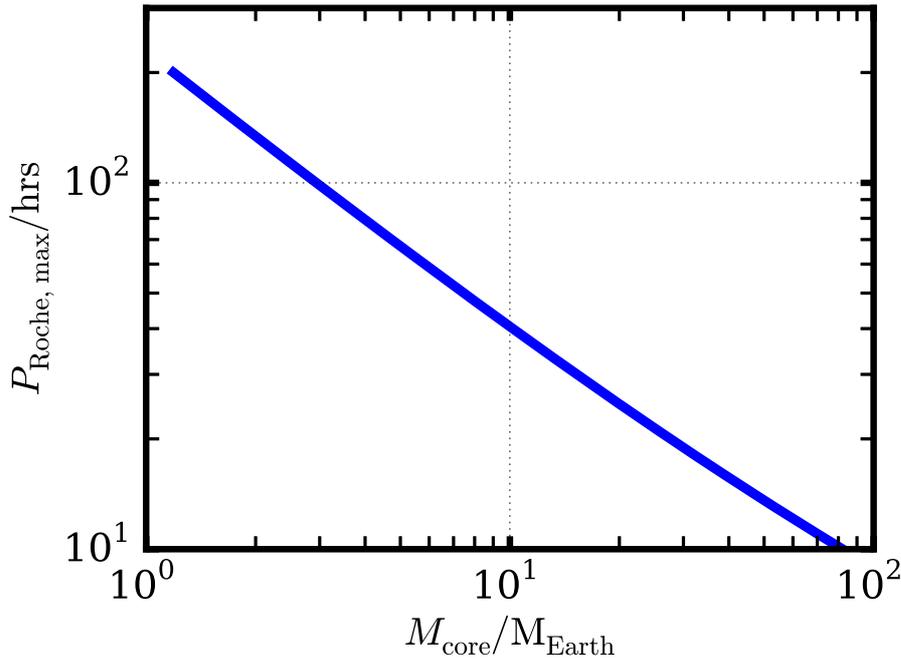}
\caption{The maximum value for the Roche limit of overflow gas giant, $P_{\rm Roche, max}$, vs. its core mass, $M_{\rm core}$.}
\label{fig:PRoche-max_vs_Mcore}
\end{figure}

Using Figure \ref{fig:LopezFortney2014_density-PRoche_contours}, we can study the orbital evolution during RLO of a gaseous planet and predict quantitatively in what orbit the orbital expansion should slow. For example, consider a gas giant planet ($f_{\rm env} \approx 1$) with $M_{\rm core} = 30\ {\rm M_{Earth}} \approx 10^{1.5}\ {\rm M_{Earth}}$. With a density like Jupiter's, 1 g/cm$^3$, such a planet would encounter its Roche limit at $P_{\rm Roche} =$ 9 hours. Overflow would begin, reducing $f_{\rm env}$ and driving the planet downward along constant $M_{\rm core}$ in the figure. The planet should eventually encounter the contour with $\rho_{\rm p} = 0.25\ {\rm g/cm^3}$ and $P_{\rm Roche} = 19\ {\rm hrs}$ contour. As $f_{\rm env} \rightarrow 0.5 \approx 10^{-0.3}$, that $\rho_{\rm p}$--$P_{\rm Roche}$ contour turns over, $\rho_{\rm p}$ increases while $P_{\rm Roche}$ decreases, and presumably the planet would either move back in toward to the star or would be stranded near this $P_{\rm Roche, max} = 19\ {\rm hrs}$. The rest of the atmosphere may be shed by photoevaporation, depending on the rate of evaporative mass loss. In any case, for this simple scenario, a planet with a known $M_{\rm core}$ (and which may or may not retain an atmosphere) should have a period near or interior to the extremum point in contour that passes nearest its $M_{\rm core}$-value. Based on the contours in Figure \ref{fig:LopezFortney2014_density-PRoche_contours}, Figure \ref{fig:PRoche-max_vs_Mcore} shows $P_{\rm Roche, max}$ as a function of $M_{\rm core}$. 

To make that comparison for known planets, consider first Kepler-78 b, represented by the green, dotted lines in Figure \ref{fig:LopezFortney2014_density-PRoche_contours}. With $P = 8.5$ hrs, Kepler-78 b has a mass (1.69 ${\rm M_{Earth}}$) and radius (1.20 ${\rm R_{Earth}}$) consistent with a bare, rocky planet, i.e. $M_{\rm core} =$ 1.69 ${\rm M_{Earth}}$. The extremum point in the green, dotted contour in Figure \ref{fig:LopezFortney2014_density-PRoche_contours} suggests such a small orbital period would require a gas giant with $M_{\rm core} > 100\ {\rm M_{Earth}}$, much larger than Kepler-78 b's current mass. Conversely, if it were the remnant of a gaseous planet, we would expect Kepler-78 b to have a period nearer to 150 hrs, much larger than observed. In principle, tidal decay of the core could reduce the orbital period, but moving from 150 to 8.5 hrs even with $Q_\star = 10^5$ would require 1.3 trillion years, and $Q_\star$-values small enough to do the job in a short enough time would give the planet a pending lifetime short enough to be statistically unlikely. This result is consistent with \cite{2015ApJ...813..101V}, which suggested that Kepler-78 b, like many other ultra-short period planets \cite{2014ApJ...787...47S}, is too close-in to be the remnant of a gas giant.

Another illustrative case is GJ 1214 b, with $P =$ 37.92 hrs. Based on its mass (6.46 ${\rm M_{Earth}}$) and radius (2.67 ${\rm R_{Earth}}$), \cite{Lopez2014Understanding} estimate $f_{\rm env} =$ 3.83\%. (Considering a different fiducial age does not significantly modify this value.) The extremum point for the $P_{\rm Roche} =$ 37.92 hrs contour lies near $M_{\rm core} =$ 10 ${\rm M_{Earth}}$, slightly larger than GJ 1214 b's core mass. Said another way, the remnant of a gas giant with a core having GJ 1214 b's core mass would be expected near or interior to $P =$ 55.6 hrs. Tidal decay of the remnant's orbit could move such a remnant inward to GJ 1214 b's current orbit, but with its current mass and $Q_\star = 10^5$, moving GJ 1214 b from 55.6 to 39.92 hours would require more than 60 Gyrs. A smaller $Q_\star = 10^4$ would bring the decay timescale down to a value closer to the system's current age (6 Gyrs) and would leave almost 1.5 Gyrs before the planet decayed down to its current Roche limit at $P_{\rm Roche} = 7$ hrs. As an M-dwarf, GJ 1214 is expected to have a deep convective zone, but it is not clear that such small tidal dissipation parameter obtains even in the case of an M-dwarf.

Finally, consider Kepler-21 b, with $P =$ 66 hrs. Based on its mass (3.85 ${\rm M_{Earth}}$) and radius (1.64 ${\rm R_{Earth}}$), we estimate its $f_{\rm env} \le$ 0.04\%. Turning to Figures \ref{fig:LopezFortney2014_density-PRoche_contours} and \ref{fig:PRoche-max_vs_Mcore}, we see that core mass puts the planet not too far from the corresponding $P_{\rm Roche, max}$, 82 hrs. In fact, tidal decay with the planet's current mass and $Q_\star = 10^5$ can move the planet from 82 to 66 hrs in about 2 Gyrs, leaving 1.4 Gyrs before the planet encounters its current Roche limit at 4.4 hrs. The system's age is not known, but these timescales are qualitatively consistent with our intuitive expectations that we should not have caught the system in a short-lived phase of its life.

\begin{figure}
\includegraphics[width=\textwidth]{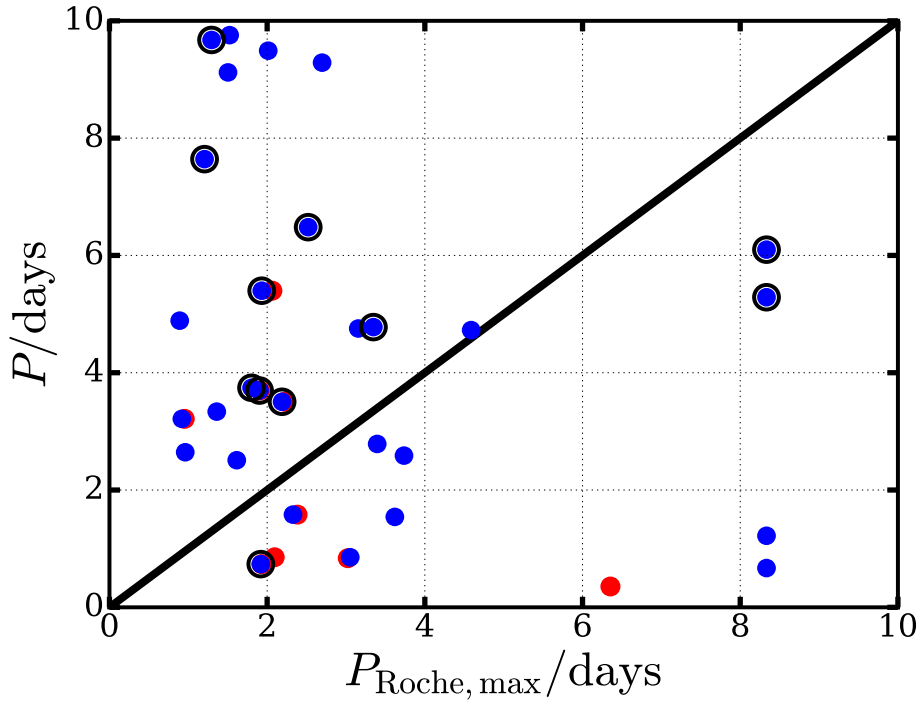}
\caption{Based on data harvested from exoplanets.org on 2016 Feb 11, observed orbital periods $P$ and estimated $P_{\rm Roche, max}$-values for many short-period planets with estimated envelope mass fractions $f_{\rm env} \le 0.1$. The blue dots represent estimates from this study, while red dots those from \cite{Lopez2014Understanding}. Black circles show which planets are members of multiplanet systems.}
\label{fig:P_vs_PRoche-max}
\end{figure}

Figure \ref{fig:P_vs_PRoche-max} compares the observed orbital periods for close-in planets to our estimates of their $P_{\rm Roche, max}$-values, shown as blue circles, and we've also considered the estimates of $f_{\rm env}$ and $M_{\rm core}$ from \cite{Lopez2014Understanding}. Here we've only considered those planets with estimated $f_{\rm env} \le 0.1$, but we have included planets (indicated with black circles) that are observed to have sibling planets. It's not clear that the multiplanet system would remain dynamically stable with one of the planets undergoing RLO, but considering that issue is outside the scope of this paper.

Figure \ref{fig:P_vs_PRoche-max} does not show a clear clustering of planets along the $P = P_{\rm Roche, max}$ line, but several objects lie nearby, qualitatively consistent with the RLO picture outlined here. Interestingly, the planet lying nearest the line is Kepler-93 b, with the most precisely measured exoplanet radius \cite{2014ApJ...790...12B}. The majority of objects depicted lie above the line, however. Some of the objects also lie well below the line, including Kepler-78 b (the red dot near $P_{\rm Roche, max} =$ 6 days) and might require implausibly small $Q_\star$-values to be consistent with the RLO models presented here. A more detailed analysis, tailored to individual systems and incorporating observational biases would clarify the origins of these objects but is beyond the scope of this study. (An interactive version of Figure \ref{fig:P_vs_PRoche-max} is available at http://www.astrojack.com/research.)

\section{Discussion and Conclusions}
\label{sec:Discussion_and_Conclusions}
By investigating the evolution of bulk density for an overflowing gas giant, we have constructed a simple model of orbital evolution during Roche lobe overflow (RLO) and sought remnant cores amongst the observed population of short-period planets. We have confirmed the results of \cite{2015ApJ...813..101V} that the degree of orbital expansion accompanying RLO is directly related to the mass of remaining planet. We have also expanded those results to argue that the core mass of the overflowing gas giant dominates the maximum orbital period reached during RLO, with the initial planetary mass, orbit, and the rates of orbital evolution and mass transfer playing important but secondary roles. This approach has the benefit of providing specific observational predictions.

As useful as our analysis here may be, it involves several important approximations and assumptions. Foremost, we have assumed the mass transfer is stable. However, unstable mass transfer may be quite important for some systems. \cite{2012MNRAS.425.2778M} pointed out that, for planets with densities in a specific range, the Roche limit may lie very near but still outside the host star. In these cases, significant angular momentum may be lost from the orbit since the innermost edge of the accretion disk orbits very near the host star's surface. If it applies, such a scenario may account for the emerging population of very short period planets. 

We have also not explicitly included the photoevaporative mass loss, which may be a key process in reducing the Neptune/sub-Neptune remnant of a hot Jupiter nearly shorn of its atmosphere to a completely denuded solid core. As discussed in Section \ref{sec:dynamics_of_planetary_rochelobe_overflow}, models for this process have been developed previously, but those models may only be applicable when a planet is not in Roche-lobe contact. However, the details of this transition probably does not significantly modify the orbital evolution considered here.

We have also assumed a solar metallicity for the planetary atmospheres. Enhanced metallicity can reduce the radius for a given mass, but the resulting enhanced atmospheric opacity can also slow the internal cooling of the planet, resulting in a slightly inflated radius at a given age \cite{2007ApJ...661..502B}. We leave an exploration of metallicity's role on RLO for future work.

Our modeling suggests four distinct scenarios for overflowing gas giants (Section \ref{sec:dynamics_of_planetary_rochelobe_overflow}): (1) Little to No Tidal Decay, for which the currently observed properties of short-period planets essentially reflects their initial conditions; (2) Complete Accretion of the Planet, for which RLO and tidal decay would very quickly remove short-period planets, masking much of the observational evidence for the process; (3) Incomplete RLO of the Planet's Atmosphere, in which case RLO proceeds relatively slowly and would leave behind gas-rich planets at their current Roche limits, a population not obvious among the observed planets (Figure \ref{fig:P-PRoche}); and (4) Complete RLO of the Atmosphere Alone, for which mass transfer removes the atmosphere of a hot Jupiter but leaves behind a denuded or gas-rich remnant near or interior to the maximum Roche limit attained during RLO, a population which is hinted at among the observed population (Figure \ref{fig:P_vs_PRoche-max}). 

Additional work is clearly needed to investigate the likely complicated transition between the endmember cases of RLO and photoevaporation to understand its influence on Neptunes/sub-Neptunes. RLO models tailored to individual observed systems may also prove fruitful and lend additional credence to the evolution discussed here. A better understanding of how short-period planets become short-period in the first place would also significantly help clarify the extent to which subsequent RLO played a role since distinguishing remnants of RLO from those that superficially resemble remnants is difficult. Our work here also provides additional impetus to better understand the mass-radius relationship for the puzzling population of sub-Neptunes/super-Earths that seem so common in the Milky Way \cite{2014PNAS..11112655M}. 

If additional work can show that many small, short-period planets are the remnants of gas giants, they can provide unprecedented insight in the natures of gas giant cores. For instance, the threshold core mass required to initiate gravitational accretion of a gas giant is probably $\gtrsim 10\ {\rm M_{Earth}}$ \cite{2014A&A...572A.107L} but remains poorly constrained by observation. Even among gas giants in our own solar system, estimates of the core masses for Jupiter and Saturn span a wide range -- Jupiter may not even have a solid core \cite{2010SSRv..152..423F} -- but the upcoming Juno mission will clarify the situation, at least for our solar system. Whether these constraints will directly bear on other planetary systems is unclear since these other systems likely had different formation conditions. 

One issue not explored here is the influence of stellar properties. Since the tidal interaction strength depends very sensitively on stellar radius (we only considered solar mass stars here), we would expect RLO of hot Jupiters to be more common for larger stars, all other things being equal. Of course, all other things are not equal, and the occurrence rate and initial conditions for hot Jupiters for different stellar types should both play key roles. The occurrence rate of RLO remnant planets should depend on the occurrence rates of hot Jupiters, but the dependence of that latter rate on stellar properties is poorly known \cite{2015ApJ...799..229W}. By contrast, \cite{2013ApJ...767...95D} reported that the occurrence rate for Earth-size (0.5-1.4 ${\rm R_{Earth}}$) planets with orbital periods shorter than 50 days is constant among cool ($T <$ 4,000 K) stars, while the same occurrence rate for 1.4-4 ${\rm R_{Earth}}$ planets declines at cooler temperatures. Exploring all these issues in the context of the RLO hypothesis may prove fruitful.

\begin{acknowledgements}
We thank the editors for the invitation to contribute to this issue and their patience as we prepared this study. We also thank an anonymous referee and Prof.\ Ferraz-Mello for thoughtful input.
\end{acknowledgements}

\bibliographystyle{spmpsci}      
\bibliography{CeMDA_tidal-decay}   

%
%

\end{document}